\documentclass[]{aa}
\usepackage{graphicx}

\begin{document}
\title{ RR Lyrae stars in the southern globular cluster NGC 362}
\author{P. Sz\'ekely \inst{1,2}
\and L. L. Kiss \inst{3}
\and R. Jackson\inst{3}
\and A. Derekas\inst{3}
\and B. Cs\'ak \inst{4}
\and K. Szatm\'ary\inst{1}}

\institute{Department of Experimental Physics \& Astronomical Observatory, University
of Szeged, H-6720 Szeged, D\'om t\'er 9., Hungary
\and Hungarian E\"otv\"os Fellowship, School of Physics, University of Sydney, NSW 2006
Australia
\and School of Physics, University of Sydney 2006, NSW 2006 Australia
\and Department of Optics and Quantum Electronics, University of Szeged, H-6720 Szeged, 
D\'om t\'er 9., Hungary}

\offprints{P. Sz\'ekely,\\
\email{pierre@physx.u-szeged.hu}}
\date{}
\abstract{}
{NGC~362 is a bright southern globular cluster in the foreground of the Small Magellanic
Cloud (SMC), for which no extensive variability survey has ever been done. Finding
regularly pulsating RR~Lyrae stars in the cluster can lead to improved metallicity and
distance estimates of the system, while other types of variable objects may be used to
confirm the results.}
{Time-series CCD photometric observations have been obtained. Light curves have been
derived with both profile fitting photometry and image subtraction. We developed a simple
method to convert flux phase curves to magnitudes, which allows the use of empirical
light curve shape vs. physical parameters calibrations.    Periods and light curve
parameters of the detected variable stars  have been determined with Fourier analysis,
phase dispersion minimization and  string-length minimization. Using the RR~Lyrae 
metallicity and luminosity calibrations, we have determined the relative  iron
abundances and absolute magnitudes of the stars. The color-magnitude diagram has been
fitted with Yale-Yonsei isochrones to determine reddening and distance independently.
For five RR~Lyrae stars we obtained radial velocity measurements from optical spectra.}
{We found 45 RR~Lyr stars, of which the majority are new discoveries. While
most of them are cluster members, as shown by their radial velocities and 
positions in the color-magnitude
diagram, we also see a few stars in the galactic field and in
the outskirts of the SMC. About half of the RR~Lyraes exhibit light curve changes
(Blazhko effect). The RR~Lyrae-based metallicity of the cluster is
$[Fe/H]=-1.16\pm0.25$, the mean absolute magnitude of the RR~Lyrae stars is $M_{\rm
V}=0.82\pm0.04$ mag implying a distance of 7.9$\pm$0.6 kpc. The mean period of RRab
stars is 0.585$\pm$0.081 days. These properties place NGC~362 among the
Oosterhoff type I globular clusters. The isochrone fit implies a slightly larger 
distance of 9.2$\pm$0.5 kpc and an age of 11$\pm$1 Gyr.
We also found 11 eclipsing binaries, 14 pulsating stars of other types, including 
classical Cepheids in the SMC and 15 variable stars with no firm classification.}
{NGC~362 hosts a large number of RR~Lyrae stars, which makes the cluster 
a potentially important test object for studying the Blazhko effect in a chemically
homogeneous environment.}


\maketitle

\section{Introduction}

Although globular clusters (GCs) play an important role in testing stellar 
evolutionary models, one can still find suprisingly bright neglected GCs,
especially in the southern hemisphere. NGC~362 is such a cluster, having been target
of  relatively few studies and so far lacking any comprehensive search for variable
stars, particularly in the central regions of the cluster. Apart from the pioneering
work  of Sawyer (1931), who presented data for fourteen variable stars, recent works
mainly aimed at obtaining deep Color-Magnitude Diagrams (CMDs, e.g. Alcaino 1976;
Harris 1982; Bolte 1986, 1987; Bellazzini et al. 2001). In the catalog of globular
cluster variables (Clement et al. 2001,  Clement 2002) there are only 16 stars
from NGC~362 and many are far from the cluster  center and have ambiguous properties.

We have been carrying out a CCD photometric survey of southern GCs since
mid-2003. In this paper we discuss the results for NGC~362, which is located in front
of the outer edge of the Small Magellanic Cloud (SMC). NGC 362~lies at the position
$\alpha=01^{\rm h}03^{\rm m}14^{\rm s}$, $\delta=-70\degr50\arcmin53\arcsec$ (J2000.0),
$l= 301\degr5$, $b=-46\degr2$, located at 8.5 kpc from the Sun and $\sim$9.4 kpc
from the galactic center; the metallicity ($[Fe/H]$) is around $-1.16$, while
the Horizontal Branch is at $V_{\rm HB}$ = 15.44. The reddening is almost negligible,
$E(B-V)=0.05$ (Harris 1996).

Models for globular clusters suggest that all the member stars were formed from the same
gas cloud at approximately the same time. Physical parameters of member stars should
therefore be representative of the cluster itself. Recently, many empirical 
relations have been
calibrated for RR~Lyrae stars (Jurcsik \& Kov\'acs 1996; Kov\'acs \& Jurcsik 1996,
Jurcsik 1998, Kov\'acs \& Walker 2001), which can be used to determine metallicities,
absolute magnitudes, reddenings and other physical parameters from the light curve shapes
of these variables. Here we present an analysis of RR~Lyrae stars in 
NGC~362 and briefly discuss properties of variable stars of other types.

\section{Observations and data analysis}

\begin{table}
\caption{Log of time series observations.}
\label{logp}
\centering
\begin{tabular} {lccc}
\hline
\hline
\noalign{\smallskip}
Date & Filter & Data points & Length [h] \\
\hline
\noalign{\smallskip}
2003 July 29 & {\it V} & 15 & 1.20 \\
2003 August 2 &{\it V}& 65 & 4.33 \\
2003 August 3 &{\it V}& 42 & 2.80 \\
2003 August 4 &{\it V}& 57 & 3.80 \\
2003 August 5 &{\it V}& 42 & 2.80 \\
2003 August 7 &{\it V}& 56 & 3.73 \\
2003 August 8 &{\it V}& 25 & 1.67 \\
2003 August 9 &{\it V}& 32 & 2.13 \\
2003 October 8 &{\it V}& 31 & 2.06 \\
2003 October 10 &{\it V}& 17 & 1.13 \\
2003 October 13	&{\it V}& 52 & 3.46 \\
2003 October 14	&{\it V}& 59 & 3.93 \\
2003 October 16 &{\it V}& 64 & 4.26 \\
2003 October 17	&{\it V}& 20 & 1.33 \\
2003 October 18 &{\it V}& 19 & 1.26 \\
2004 October 2 &{\it V}& 115 & 6.85 \\
2004 October 5 &{\it V}& 106 & 7.46 \\
\hline
\end{tabular}
\begin{list}{}{}
\item Landolt fields in $V$ and $I$: 2003 August 1 \\
\end{list}
\end{table}
\begin{table}
\caption{Log of spectroscopic observations.}
\label{logs}
\centering
\begin{tabular} {lccr}
\hline
\hline
\noalign{\smallskip}
Date &  No. of spectra & Stars\\
\hline
\noalign{\smallskip}
2004 October 25 &  9, 1, 10 & V7, V27, V74 \\
2004 October 26 &  1, 2, 1 & V7, V23, V74 \\
2004 October 27 & 1 & V7 \\
2004 October 28 &  17 & V7 \\
2004 October 29 &  12, 9 & V23, V71 \\
2004 October 30 &  7 & V27 \\
2004 November 1 &  6, 10 & V27, V74 \\
2004 November 2 &  1, 4 & V7, V27 \\
\hline
\end{tabular}
\end{table}

Time-series $V$-band CCD photometric observations were carried out on 17 nights
between  July 2003 and October 2004, using the 1m ANU telescope at Siding Spring
Observatory,  Australia. For imaging we used three chips of the Wide Field Imager
giving $\sim40^\prime \times26^\prime$ field of view. In addition to the photometry, we
measured radial velocities for 5 RR~Lyrae stars from optical spectra taken with the
2.3m ANU telescope and the Double-Beam Spectrograph. We used only the red arm of the
spectrograph with the 1200 mm$^{-1}$ grating, covering the wavelength range
5800--6800 \AA. The nominal resolving power was $\lambda/\Delta \lambda\approx7000$.
With 20-min exposures, the spectra had S/N ratios of 20-50, depending on the
weather conditions.  The full journals of observations are summarized in Tables\
\ref{logp}-\ref{logs}.  

All data were processed with standard IRAF\footnote{IRAF is distributed by the National
Optical Astronomy Observatories, which are operated by the Association of Universities
for Research in Astronomy, Inc., under cooperative agreement with the National Science
Foundation.} routines, including bias and flat corrections. Cluster images were then
analysed both with PSF photometry, using the {\tt daophot} package in IRAF, and image
subtraction photometry, using the ISIS 2.1 package of Alard (2000). For standard
photometric calibrations we observed selected equatorial fields of Landolt (1992): SA
109-954, SA 110-503, SA 111-1965,  Mark A and SA 113-241. The transformation
coefficients were determined with the task $photcal$. Radial velocities were measured
with cross-correlation, using spectra of the IAU radial velocity standard HD~187691 and
the task $fxcor$, leading to an estimated accuracy of $\pm$7 km~s$^{-1}$.

Light curves from PSF-fitting and image subtraction were analysed for finding
periods. For this, we applied a combination of Fourier analysis, phase
dispersion minimization (Stellingwerf, 1978) and string-length minimization (Lafler \&
Kinman, 1965). The latter gives better results when the light curve contains sharp
features, such as narrow minima of Algol-type eclipsing binaries. Every light curve was
then inspected visually and classified according to its period, amplitude and shape.
Variable star designations were assigned in an increasing order of right ascension. 

\begin{figure}
\begin{center}
\leavevmode
\includegraphics[width=8cm]{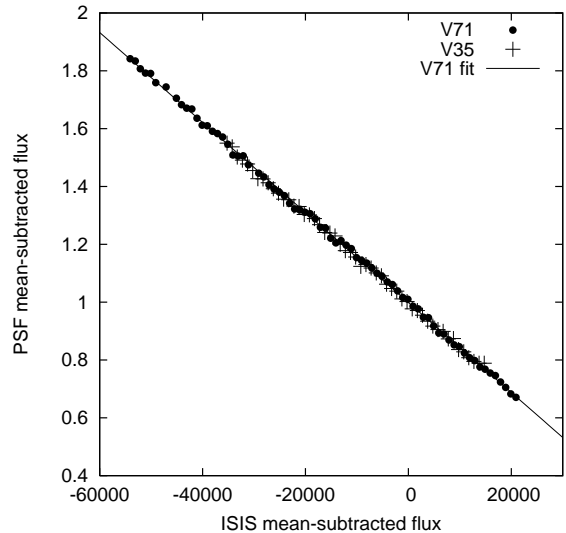}
\caption{The conversion between mean-subtracted flux values.}
\label{flux2mag}
\end{center}
\end{figure}

\begin{figure}
\begin{center}
\leavevmode
\includegraphics[width=8cm]{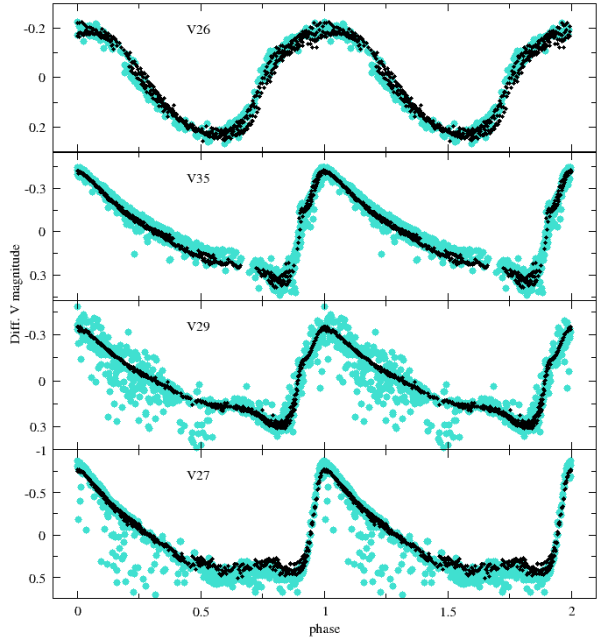}
\caption{Mean-subtracted magnitude curves obtained using {\tt daophot} (light
grey/turqoise points) and Eq.\ (1) applied to ISIS fluxes (black dots). The large
scatter in V29 and V27 is caused by crowding.}
\label{magcomp}
\end{center}
\end{figure}

Image subtraction is far more efficient than PSF photometry in detecting variable 
stars in crowded fields. Globular cluster studies are heavily affected by this, where large
incompleteness factors can arise from the highly concentrated distributions of stars
(Kaluzny et al. 2001). Even for the best-studied clusters, image subtraction can
yield new discoveries (e.g. Clementini et al. 2004). There is, however, a major
practical difficulty in using image subtraction results for the empirical light curve
shape vs. physical parameters calibrations. Whereas ISIS gives time variations of
brightness expressed in differential fluxes (relative to a master CCD frame), all the
empirical relations were calibrated with light curves in magnitudes, which have a
different shape, especially for the largest-amplitude objects. To convert fluxes to
magnitudes, one has to determine a reliable magnitude estimate for every variable star
in the master CCD frame, otherwise the conversion (and employing the calibrations) is
not possible (see, e.g., Corwin et al. 2006). 

Since we detected about twice as many RR~Lyrae stars with ISIS 2.1 than with {\tt
daophot}, we developed a simple method for a flux-to-magnitude conversion. The main
assumption is that there is a unique scaling between the mean-subtracted ISIS fluxes and
flux values calculated from the mean-subtracted {\tt daophot} magnitudes, so that one can
write the following equation:

\begin{equation}
10^{-0.4(m-\langle m\rangle)} = 1 + c(f-\langle f\rangle).
\end{equation}

\noindent Here, $f$ is the differential flux, $m$ is the differential magnitude
relative to a set of ensemble comparison stars, while $\langle \rangle$ refers to mean
values, given by the population means $\langle f \rangle=\int_0^1f(\varphi)d\varphi$ 
and $\langle m \rangle=-2.5\log(\int_0^110^{-0.4m(\varphi)}d\varphi)$ for phased fluxes
and magnitudes, respectively. Note that by writing Eq.\ (1) we also assume that the 
two population means refer to the same state of the star, which is the reason why we 
defined the mean magnitude $\langle m \rangle$ with the given expression. In principle,
the conversion factor $c$ can be determined from ISIS flux and {\tt daophot} magnitude
curves of variable stars in the outskirts of the cluster, where crowding has no effect.
Eq.\ (1) can then be applied to those variables which have only flux curves to
calculate the left-hand side of Eq.\ (1), whose logarithm multiplied by $-2.5$ gives
the magnitude curve with zero mean.

In Fig.\ \ref{flux2mag} we plot data for two high-amplitude RR~Lyrae stars which are
located in the outer regions of the field. V71 is the largest-amplitude variable with
no crowding, so its data were used to determine $c$.  The solid dots in Fig.\
\ref{flux2mag} yield a slope of $c=(-1.550\pm0.005)\times10^{-5}$. Since these
calibration points cover the whole range of ISIS fluxes for all the RR~Lyraes, no
extrapolation was needed for the other stars. For comparison, we also show data for
another uncrowded star, V35. The two sets do not differ by more than 1\% anywhere,  so
that Eq.\ 1, calibrated by V71 alone, can predict magnitudes from fluxes  within 0.01
mag (without the zeropoint). For a few other RR~Lyrae stars we also compared the
magnitude curves that were calculated from the ISIS flux data to the {\tt daophot}
magnitudes and always found excellent agreement (Fig.\ \ref{magcomp}), thus 
confirming the applicability of our approach. We therefore converted ISIS flux
curves  to magnitudes for every star with good phase coverage.    

All observations presented in this paper are available electronically at the CDS via
anonymous ftp to {\tt cdsarc.u-strasbg.fr (130.79.128.5)} or via {\tt
http://cdsweb.u-strasbg.fr/cgi-bin/qcat?J/A+A/\\.../...}.

\section{Regular RR Lyr stars}

\begin{figure}
\begin{center}
\leavevmode
\includegraphics[width=8cm]{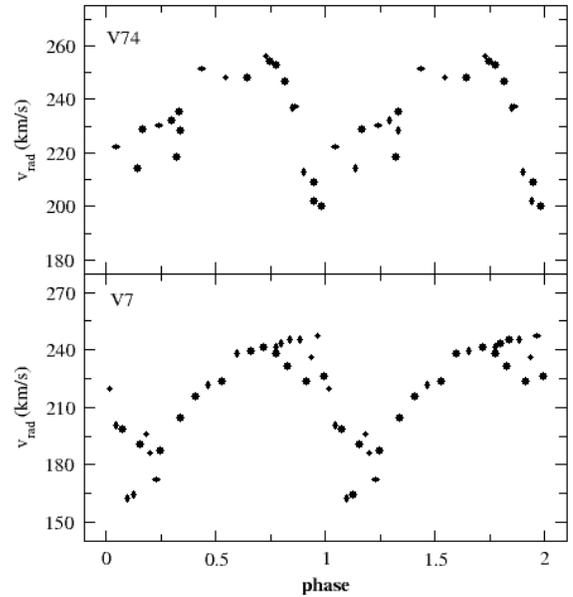}
\caption{Radial velocity curves for two RR~Lyrae stars.}
\label{rv}
\end{center}
\end{figure}

\begin{figure*}
\begin{center}
\leavevmode
\includegraphics[width=18cm]{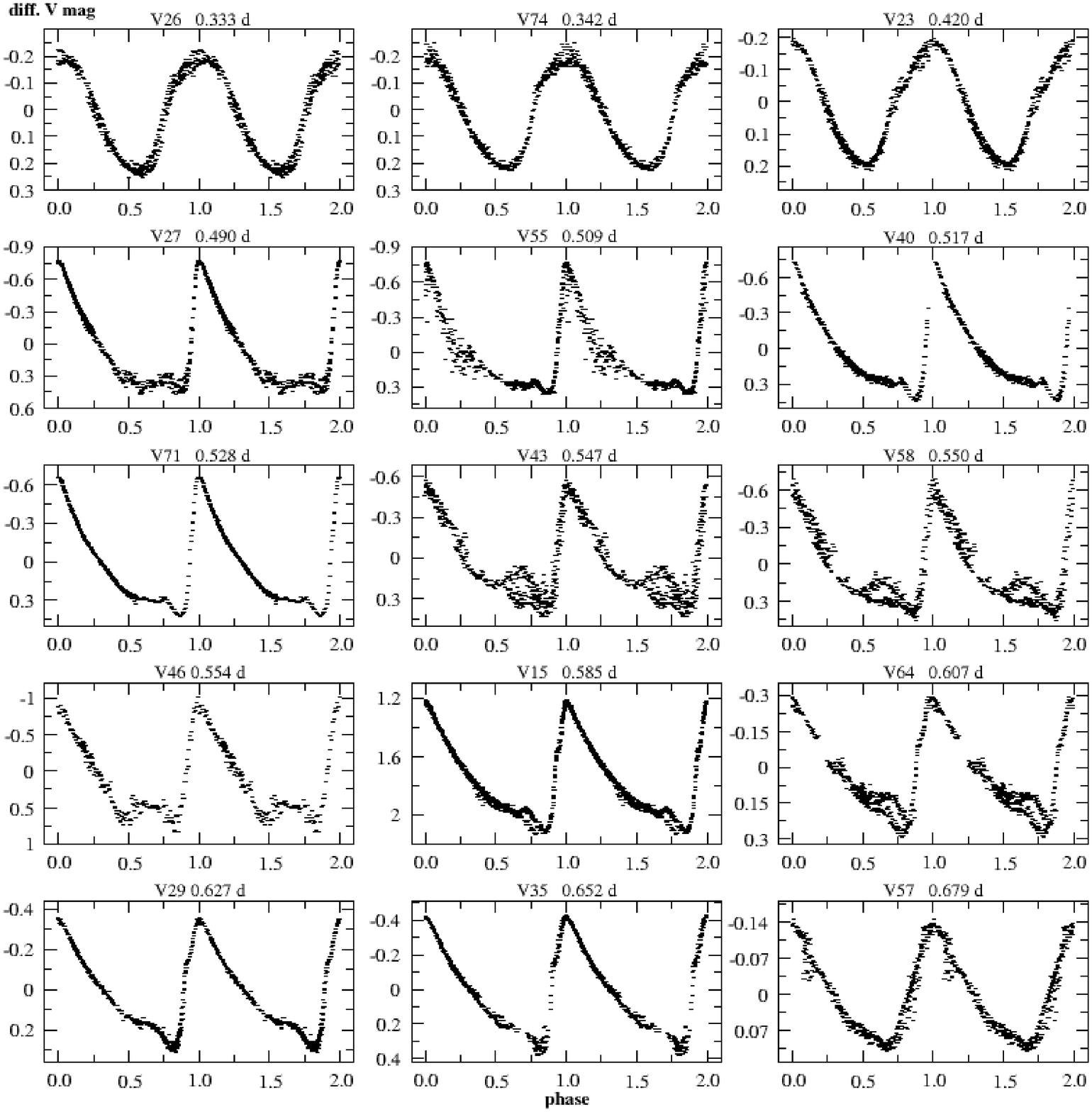}
\caption{Sample phased light curves of RR~Lyr stars, in order of increasing
pulsational period from V26 (P=0.333d) to V57 (P=0.679d).}
\label{norms}
\end{center}
\end{figure*}

\begin{figure*}
\begin{center}
\leavevmode
\includegraphics[width=18cm]{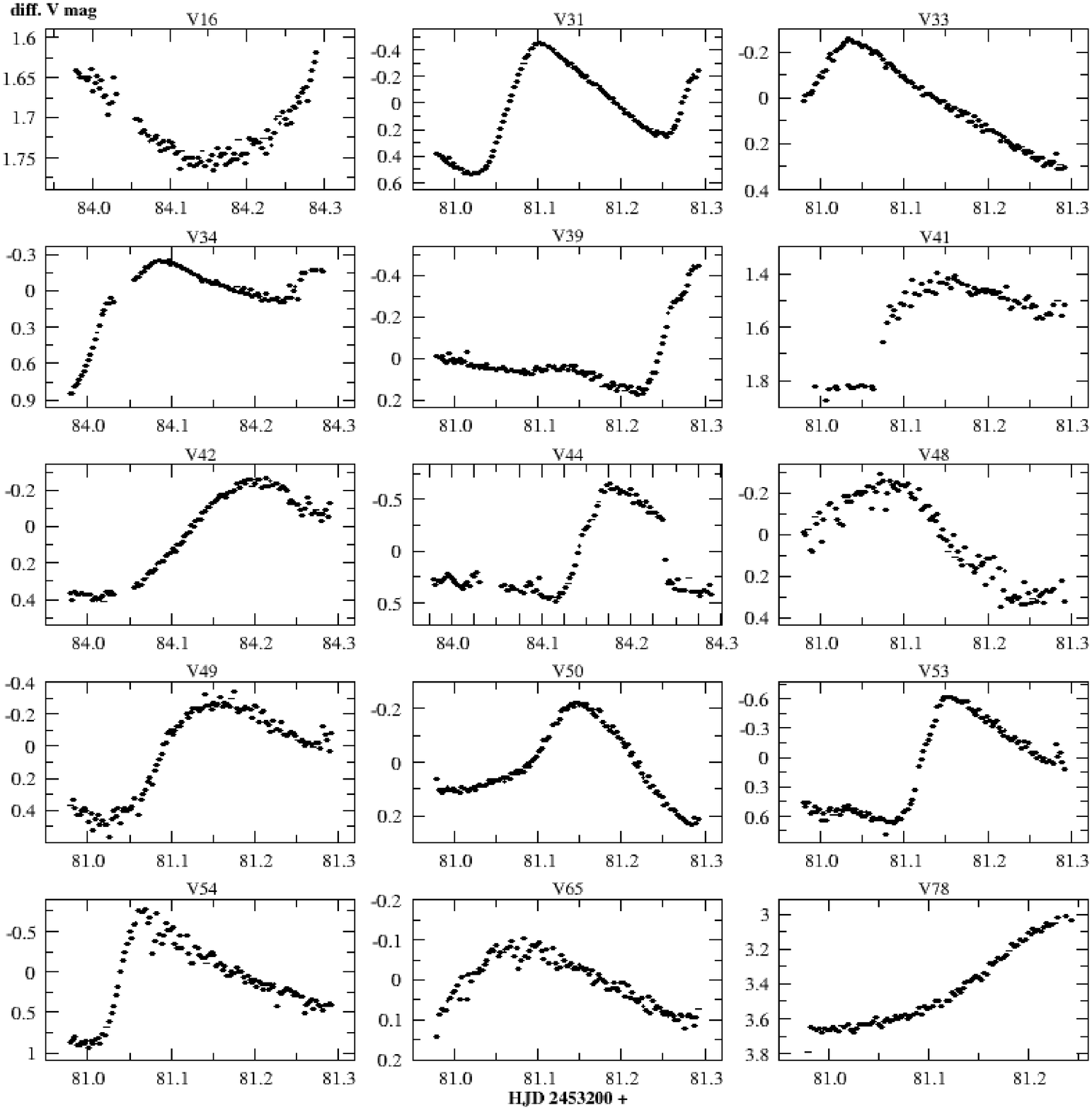}
\caption{Sample light curves of cluster RR~Lyr stars.}
\label{norms2}
\end{center}
\end{figure*}

From the periods and light curve shapes we identified 45 RRab/RRc-type variable stars. In
some cases only partial phase coverage was obtained and only the RR~Lyrae-class  could be
safely concluded. The majority have very similar mean apparent brightnesses
($V\approx15.5$ mag). Five of these have been observed spectroscopically (see Fig.\
\ref{rv} for two examples) and their mean radial velocities ($\approx$220 km~s$^{-1}$)
confirm membership, since the cluster's radial velocity is 223.5$\pm$2.0 km~s$^{-1}$
(Fischer et al. 1993). Figs.\ \ref{norms}-\ref{norms2} show sample phased/unphased
data for cluster member RR~Lyraes with well-defined mean light curve shapes and with
partial coverage, respectively. Table\ \ref{rrls} contains the fundamental photometric
data for all the detected RR~Lyrae stars. Since photometric standards were observed only
on one night, $V$ brightnesses and $V-I$ colors are subject to uncertainty due to
photometric variability. To address this, we calculated the pulsation phase of the
standard $V$ observation of the cluster for each RR~Lyrae star, so that we know from the
$V$-band light curve the difference between the magnitude in that phase and the mean. The
measured standard $V$ magnitudes were then corrected for these differences. With no
$I$-band light curve a similar correction for the standard $I$ magnitudes was not
possible, so that the $V-I$ indices may suffer from larger uncertainties (up to several
tenths of a magnitude). Stars with missing $V$ and $V-I$ in Table\ \ref{rrls} values were
not detected as individual objects in the master standard frames, which means they were
merged with other stars. Variables in the core of the cluster were checked for blending
using an archived WFPC2 image of the  Hubble Space Telescope (taken for the project ``Hot
stellar populations in globular cluster cores'',  HST Proposal 6095 by S. Djorgovski, 160
s exposure in the F439W filter). In the last column of Table\ \ref{rrls} we give comments
and identifications of previously known variables from the literature.

\begin{figure}
\begin{center}
\leavevmode
\includegraphics[width=7cm]{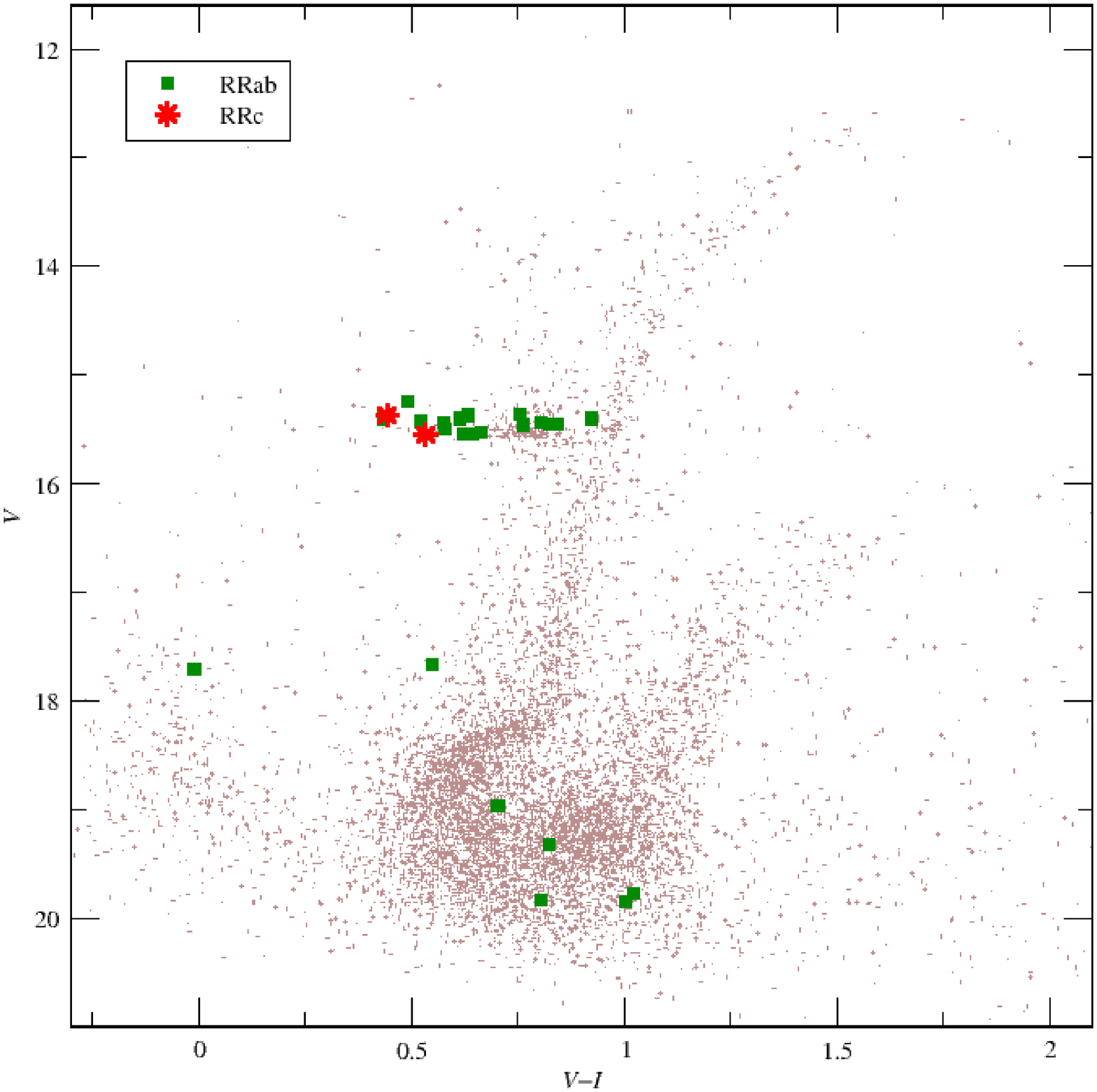}
\caption{RR~Lyrae stars in the color-magnitude diagram of NGC~362.}
\label{rrcmd}
\end{center}
\end{figure}

In Fig.\ \ref{rrcmd} we plotted the positions in the CMD of RR~Lyrae stars 
with reliable
calibrated photometry. Most of the stars fall on the
Horizontal Branch (HB) clearly establishing membership. There are
also some stars with deviant positions, for example, V80 and V81 at $V\approx17.6$,
both in the background  of the cluster with distances well beyond 20 kpc. Further
RR~Lyraes are seen between  $V=19-20$ mag, which are located in the SMC. 

\begin{figure}
\begin{center}
\hskip2mm\includegraphics[width=7.5cm]{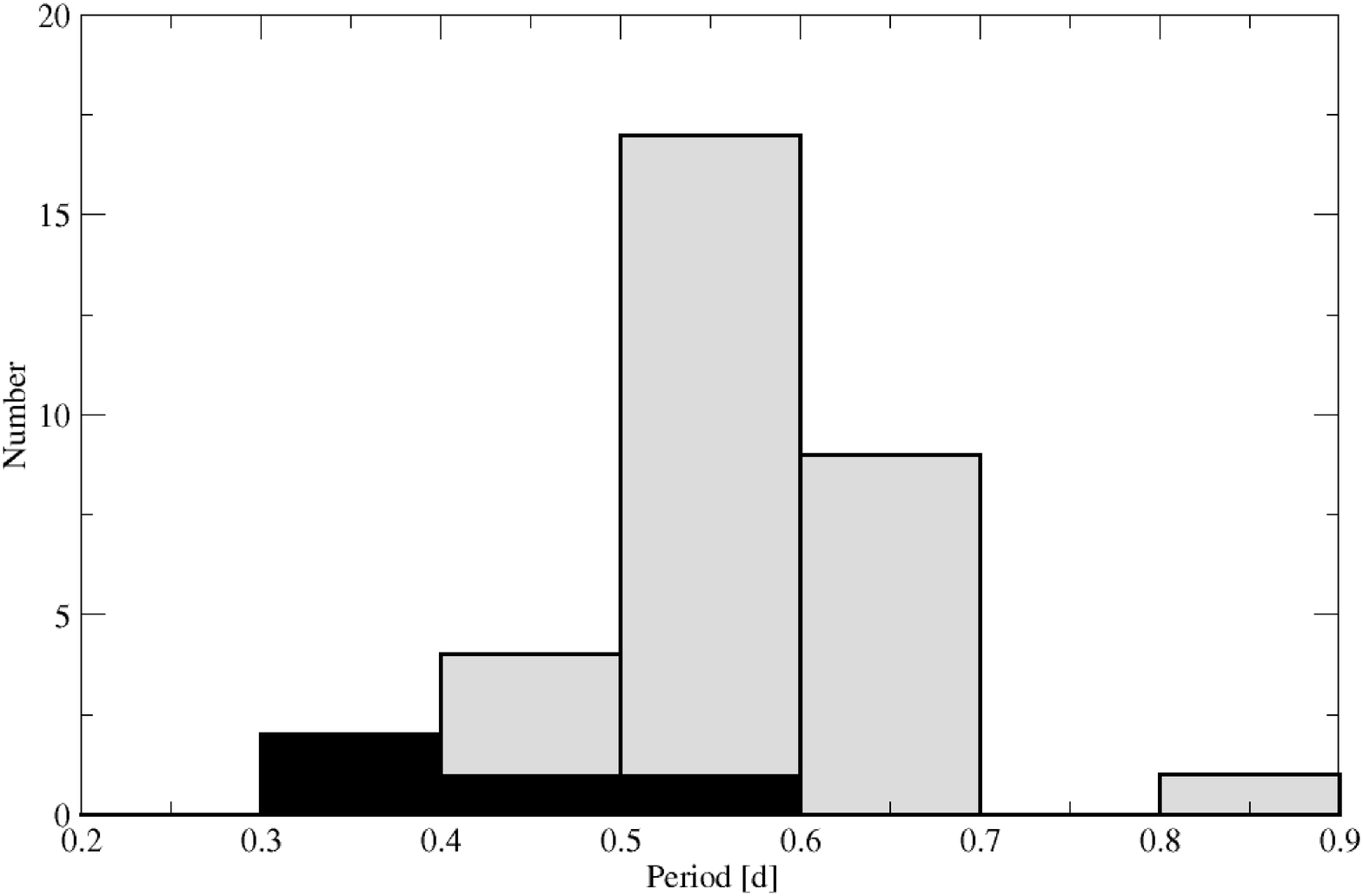}
\vskip1mm
\includegraphics[width=7.5cm]{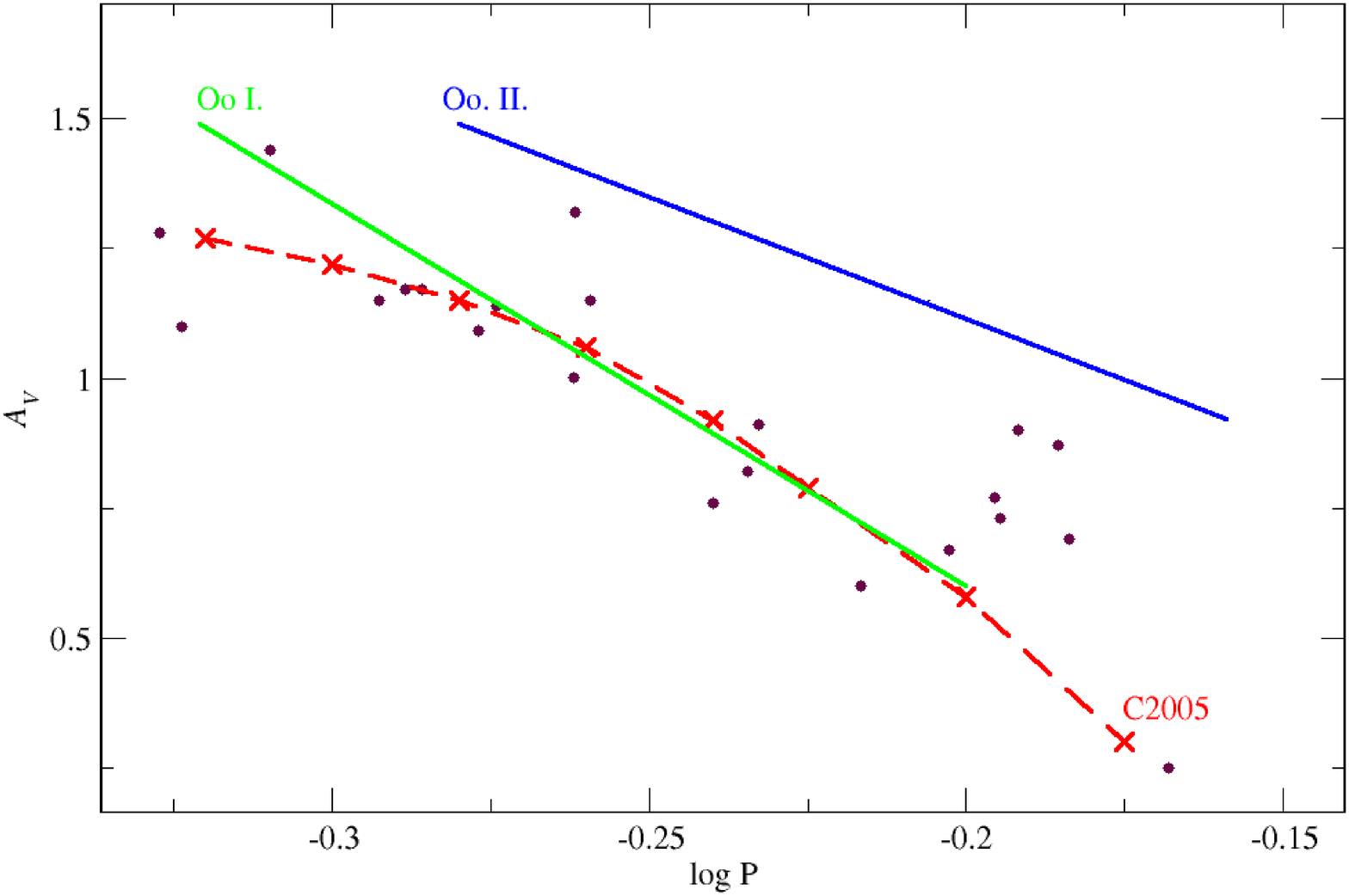}
\caption{{\it Upper panel:} Period distribution of cluster RR~Lyrae stars. 
The darker columns refer to the RRc stars. {\it Lower panel:} 
Bailey-diagram for the cluster RR~Lyrae stars. Solid lines: Clement 
\& Rowe (2000); dashed line: Cacciari et al. (2005).}
\label{rrnr}
\end{center}
\end{figure}

We could determine periods, amplitudes and epochs for 28 RR~Lyraes, of which 4 are
first-overtone (RRc) stars. This fairly low percentage implies that the cluster is
RRc-poor (cf. Fig. 1 of Castellani et al. 2003), which is characteristic for 
Oosterhoff type I globular clusters (Oosterhoff 1939). The computed specific frequency of
RR~Lyr variable stars, including the newly found ones, is $S_{\rm RR}=N_{\rm
RR}\times10^{0.4(7.5+M_V)}\approx16$. The histogram of the periods (upper panel in 
Fig.\ \ref{rrnr}) shows that large number of RRab stars have periods between 0.5 and 0.6
d. The mean period of 24 RRab stars is $\langle {\it P}_{\rm ab} \rangle = 0.585$ or
$\langle \log {\it P}_{\rm ab} \rangle = -0.237$, while the period--amplitude relation
closely  follows the Oosterhoff I lines given by Clement \& Rowe (2000) and Cacciari
et al. (2005).  Both properties are consistent with the Oosterhoff I class (see the
lower panel in Fig.\ \ref{rrnr}). 

In order to determine metallicities and absolute magnitudes of the RR~Lyrae stars, we
fitted the phase diagrams with Fourier polinomials of up to the 10th order. Using Eq. (3)
of Jurcsik and Kov\'acs (1996): $[Fe/H]=-5.038-5.394P+1.345\phi_{31}$, we estimated iron
abundances for 13 RR Lyrae stars. These objects have phase curves with low scatter that
covered both maxima and minima and did not show light curve modulations. We then
determined absolute magnitudes using the metallicity-luminosity relation of Kov\'acs \&
Jurcsik (1996):  $M_{\rm V}=0.19[Fe/H]+1.04$. Upon comparing this to our phase corrected
apparent magnitudes, we calculated distance moduli for the individual stars. We
determined other physical parameters (e.g. effective temperature, luminosity) using
further equations calibrated by Morgan et al. (2005), Kov\'acs (1998) and Jurcsik (1998).
The results are presented  in Tables\ \ref{fourab}-\ref{fourc}.

The numbers clearly show that most of the studied RR~Lyraes define reasonably small
ranges of the parameters. The distance to V58 is erroneous because the star is a blend,
but its other, distance-independent, parameters are similar to those of the other
variables.  The calculated mean metallicity of the cluster is in good agreement with values 
found in the literature, e.g. Caldwell (1988) and Davidge (2000), while the distance 
is slightly smaller, but still compatible with, the 8.5 kpc in Harris (1996). In conclusion,
RR~Lyraes in NGC~362 define a consistent set of physical parameters for the 
cluster.

\begin{table*}
\begin{scriptsize}
\caption{RR~Lyrae stars in the field of NGC~362. The given coordinates are accurate 
within $\pm0.1-0.2^{\prime\prime}$. Remarks on types: 1. strong Blazhko 
modulation, the whole light curve is changing; 2. slight modulation in certain phases; 
3. blend; 4. unsure. The last column 
contains other designations of already known variables:  SAW xxx --- Sawyer (1931); 
SMC Vxxx --- Samus et al. (2004); MA 1315 --- Meyssonnier \& Azzopardi (1993); SSP xx and HV xx --- 
Sharpee et al. (2002).}
\label{rrls}
\centering
\begin{tabular}{ccclccccll}
\hline
\hline
\noalign{\smallskip}
Designation & $\alpha$ [h : m: s] & $\delta$ [\r{} : $\arcmin$ : $\arcsec$ ] & Period [d] & $V$ & $V-I$ & Ampl. & Epoch & Type & Comments \\
&&&&&&[mag]& [HJD$-$2\,400\,000]&&\\
\hline
\noalign{\smallskip}
V1 & 1:00:56.64 & $-$71:01:43.8 & 0.52835 & & & 0.55 & 2860.1765 & RRab & halo/SMC, SSP 23\\
V6 & 1:01:49.14 & $-$71:07:36.9 & 0.30250 & 18.96 & 0.71 & 1.07 & 3284.1702 & RRab & halo/SMC \\
V7 & 1:01:54.55 & $-$70:46:59.7 & 0.54753 & 15.80 & $-$0.12 & 1.32 & 2854.2440 & RRab$^1$ & SMC V1670\\
V8 & 1:01:56.71 & $-$71:10:00.2 & 0.29218 & 19.33 & 0.83 & 0.98 & 3284.1647 & RRab & halo/SMC, SSP 29 \\
V10 & 1:01:58.12 & $-$70:50:29.8 & & & & & & RR & \\
V12 & 1:02:00.15 & $-$70:59:02.3 &        & 19.77 & 1.03 & 0.35 &            & RR   &halo/SMC \\
V15 & 1:02:24.02 & $-$70:51:56.4 & 0.58506 & 15.55 & 0.58 & 0.91 & 2927.0715 & RRab$^2$ & SAW V1, SMC V1712 \\
V16 & 1:02:24.06 & $-$71:05:13.5 & 0.59339 & 16.47 & 0.11 & 0.25 & 2927.1882 & RRc$^4$ & MA 1315 \\
V23 & 1:02:46.55 & $-$70:51:16.7 & 0.42023 & 15.37 & 0.45 & 0.41 & 3281.0256 & RRc$^2$ & SAW V10, SMC V1740 \\
V26 & 1:02:57.22 & $-$70:50:40.8 & 0.33329 & 15.07 & 1.09 & 0.47 & 3281.2486 & RRc$^2$ & \\
V27 & 1:02:58.08 & $-$70:51:23.6 & 0.49009 & 15.50 & 0.44 & 1.44 & 3281.0826 & RRab$^2$ & SAW V5, SMC V1572 \\
V29 & 1:03:01.93 & $-$70:49:43.2 & 0.62709 & 15.44 & 0.58 & 0.67 & 3284.2818 & RRab & \\
V31 & 1:03:04.29 & $-$70:51:19.8 & $\sim$0.5 & 14.62: & 0.38: & 0.6 & & RR$^3$ & SMC V1757 \\
V33 & 1:03:05.31 & $-$70:51:03.7 & 0.63895 &  &  & 0.73 & 3281.0148 & RRab$^2$ & \\
V34 & 1:03:07.23 & $-$70:50:53.4 & 0.63757 & 15.24 & 0.50 & 0.77 & 2927.0661 & RRab & \\
V35 & 1:03:07.55 & $-$70:52:49.2 & 0.65256 & 15.46 & 0.64 & 0.87 & 3284.1213 &RRab & SAW V12, SMC V1760 \\
V37 & 1:03:07.88 & $-$70:50:38.7 & & 14.48 & 0.34 & & & RR$^3$ & \\
V38 & 1:03:07.92 & $-$70:50:42.4 & & 15.45 & 0.85 & & & RR & \\
V39 & 1:03:08.06 & $-$70:50:06.8 & 0.58277 &  &  & 0.82 & 2855.2943 & RRab$^2$ & \\
V40 & 1:03:09.08 & $-$70:52:00.2 & 0.51772 & 14.96 & 0.88 & 1.17 & 3284.0559 & RRab$^2$ & SMC V1762 \\
V41 & 1:03:09.16 & $-$70:49:45.1 & & 15.44 & 0.81 & 0.54 & & RR & \\
V42 & 1:03:09.17 & $-$70:50:43.2 & 0.655 & 15.43 & 0.53 & 0.69 & 3284.2027 & RRab & \\
V43 & 1:03:11.31 & $-$70:50:25.4 & 0.54714 & 15.40 & 0.93 & 1.00 & 3281.1316 & RRab & \\
V44 & 1:03:11.62 & $-$70:50:46.8 & 0.532 & 13.60: & 1.08: & 1.14 & 3284.1753 & RRab$^3$ & \\
V45 & 1:03:11.62 & $-$70:50:51.1 & 0.503 & 14.03 & 0.62 & & & RRab & \\
V46 & 1:03:12.52 & $-$70:50:49.8 & 0.55417 & 12.98: & 0.63: & 1.85 & 3281.1178 & RRab$^3$ & \\
V48 & 1:03:13.89 & $-$70:50:53.1 & & & & & & RR & \\
V49 & 1:03:14.09 & $-$70:50:41.5 & 0.643 &  &  & 0.90 & 3281.1589 & RRab & \\
V50 & 1:03:14.50 & $-$70:51:50.7 & & 15.45: & 0.83: & & & RR$^3$ & \\
V53 & 1:03:17.19 & $-$70:50:43.4 & 0.532 & & & & & RRab & \\
V54 & 1:03:17.28 & $-$70:50:37.9 & 0.528 & & & & & RRab & \\
V55 & 1:03:17.57 & $-$70:50:15.7 & 0.50984 & 15.49 & $-$0.12 & 1.15 & 3284.1132 & RRab & SMC V1780 \\
V57 & 1:03:17.63 & $-$70:49:27.4 & 0.67901 & 15.54 & 0.65 & 0.25 & 2929.2059 & RRab & \\
V58 & 1:03:18.56 & $-$70:51:16.7 & 0.55060 & 13.40 & 1.33 & 1.15 & 3281.1316 & RRab$^1$ & \\
V62 & 1:03:21.30 & $-$70:50:35.2 & & 15.56 & 0.97 & & & RR & \\
V64 & 1:03:22.87 & $-$70:50:35.8 & 0.60726 & 15.50 & 0.76 & 0.60 & 2929.1598 & RRab & \\
V65 & 1:03:24.17 & $-$70:50:43.7 & 0.691 & & & & & RRab$^3$ & \\
V67 & 1:03:31.24 & $-$70:50:41.7 & 0.51464 & 15.52 & 0.62 & 1.17 & 3281.0691 & RRab$^1$ & SAW V6, SMC V1800 \\
V68 & 1:03:32.57 & $-$70:53:19.7 & 0.47446 & 15.60 & 1.07 & 1.10 & 3284.1810 & RRab$^1$ & SAW V3, SMC V1804 \\
V71 & 1:03:40.35 & $-$70:51:18.7 & 0.52856 & 15.54 & 0.63 & 1.09 & 3281.0038 & RRab & SAW V7, SMC V1816 \\
V73 & 1:03:50.37 & $-$70:37:51.7 &         & 19.83 & 0.81 & 0.9 & & RR & halo/SMC\\
V74 & 1:03:56.72 & $-$70:46:19.0 & 0.34249 & 15.55 & 0.54 & 0.48 & 3281.2160 & RRc & \\
V78 & 1:04:56.99 & $-$70:44:12.2 & 0.83238 &  &  & 0.72 & 2854.2549 & RRab$^2$ & \\
V80 & 1:05:14.82 & $-$71:03:10.1 & 0.47084 & 17.66 & 0.55 & 1.28 & 3281.1234 & RRab & halo \\
V81 & 1:05:33.24 & $-$71:00:43.4 & 0.57571 & 17.71 & --0.01 & 0.76 & 2921.1516 & RRab$^2$ & halo \\
\hline
\end{tabular}
\end{scriptsize}
\end{table*}

\begin{table*}
\begin{scriptsize}
\caption{Fourier and the physical parameters of selected RRab stars. Notes:  $^\dag$ Blazhko effect present. $^x$ Blend. $^*$ Outliers, not included in the cluster average}
\label{fourab}
\centering
\begin{tabular}{lcccccccccccr}
\hline
\hline
\noalign{\smallskip}
Star & $\phi_1$ & $\phi_3$ & $\phi_4$ & A$_1$ & Period [d] & $(V-I)_0$ & $(V-K)_0$ & $[Fe/H]$ & $\log T_{\rm eff}$ & $\log L/L_{\odot}$ & $M_{\rm V}$ & D [kpc] \\
\hline
\noalign{\smallskip}
V15 & 0.752 & 1.242 & 1.658 & 0.308 & 0.5851 & 0.52 & 0.59 & $-$1.11 & 3.867 & 1.58 & 0.80 & 8.4 \\
V27 & 0.828 & 0.917 & 1.247 & 0.494 & 0.4901 & 0.55 & 1.02 & $-$1.34 & 3.819 & 1.59 & 0.79 & 8.3 \\
V29 & 0.736 & 1.414 & 1.938 & 0.221 & 0.6271 & 0.51 & 0.38 & $-$1.04 & 3.890 & 1.58 & 0.80 & 7.9 \\
V35 & 0.805 & 1.689 & 5.513 & 0.267 & 0.6525 & 0.67 & -0.74 & $-$1.08 & 4.015 & 1.60 & 0.75 & 8.2 \\
V40 & 0.698 & 0.982 & 0.919 & 0.215 & 0.5177 & 0.48 & 0.67 & $-$0.88 & 3.858 & 1.53 & 0.93 & 6.0 \\
V43 & 0.795 & 0.992 & 1.410 & 0.299 & 0.5471 & 0.52 & 0.71 & $-$1.41 & 3.854 & 1.58 & 0.82 & 7.8 \\
V55 & 0.917 & 0.988 & 0.965 & 0.290 & 0.5098 & 0.54 & 0.85 & $-$1.71 & 3.840 & 1.58 & 0.83 & 8.2 \\
V58$^{x,*}$ & 0.822 & 1.182 & 1.239 & 0.335 & 0.5506 & 0.54 & 0.76 & $-$1.28 & 3.849 & 1.58 & 0.81 & 3.1 \\
V64 & 0.886 & 1.838 & $-$0.293 & 0.195 & 0.6072 & 0.62 & 0.78 & $-$0.96 & 3.845 & 1.56 & 0.84 & 8.1 \\
V67$^\dag$ & 0.606 & 0.780 & 0.981 & 0.339 & 0.5146 & 0.49 & 0.82 & $-$0.76 & 3.840 & 1.54 & 0.88 & 8.0 \\
V71 & 0.739 & 0.933 & 1.152 & 0.375 & 0.5286 & 0.53 & 0.85 & $-$1.16 & 3.838 & 1.57 & 0.82 & 8.2 \\
V80$^*$ & 0.904 & 1.073 & 1.450 & 0.427 & 0.4708 & 0.29 & 0.89 & $-$1.33 & 3.834 & 1.57 & 0.84 & 22 \\
V81$^*$ & 1.049 & 0.914 & 1.770 & 0.283 & 0.5757 & 0.40 & 0.75 & $-$2.69 & 3.854 & 1.66 & 0.70 & 24 \\
\hline
\noalign{\smallskip}
Average & 0.78 & 1.18 & 1.52 & 0.30 & 0.557 & 0.51 & 0.60 & $-$1.16 & 3.865 & 1.57 & 0.82 & 7.9 \\
$\sigma$ & 0.08 & 0.32 & 1.37 & 0.08 & 0.05 & 0.08 & 0.45 & 0.25 & 0.05 & 0.14 & 0.04 & 0.6
 \\
\hline
\end{tabular}
\end{scriptsize}
\end{table*}

\begin{table*}
\begin{scriptsize}
\caption{Fourier and physical parameters of selected RRc stars.}
\label{fourc}
\begin{center}
\begin{tabular}{lccccccccccr}
\hline
\hline
\noalign{\smallskip}
Star & $\phi_1$ & $\phi_2$ & $\phi_3$ & A$_4$ & Period [d] & $M/M_{\odot}$ & $\log L/L_{\odot}$ & $\log T_{\rm eff}$ & $\log Y$ & $M_{\rm V}$ & D [kpc] \\
\hline
\noalign{\smallskip}
V23 & 0.045 & 0.498 & 1.706 & 0.0056 & 0.4202 & 1.052 & 1.93 & 3.821 & $-$1.042 & 0.814 & 7.7 \\
V26 & $-$0.160 & 1.670 & 0.313 & 0.0083 & 0.3333 & 1.134 & 1.83 & 3.839 & $-$0.990 & 0.816 & 6.7 \\
V74 & $-$0.145 & 1.233 & 0.259 & 0.0087 & 0.3425 & 1.179 & 1.88 & 3.838 & $-$1.001 & 0.826 & 8.3 \\
\hline
\noalign{\smallskip}
Average & $-$0.086 & 1.133 & 0.757 & 0.0075 & 0.365 & 1.121 & 1.89 & 3.833 & $-$1.011 & 0.818 &  7.6 \\
$\sigma$ & 0.09 & 0.48 & 0.67 & 0.001 & 0.04 & 0.05 & 0.02 & 0.008 & 0.022 & 0.005 & 0.7 \\
\hline
\end{tabular}
\end{center}
\end{scriptsize}
\end{table*}

\section{RR~Lyr stars with Blazhko effect}

\begin{figure}
\begin{center}
\leavevmode
\includegraphics[width=8cm]{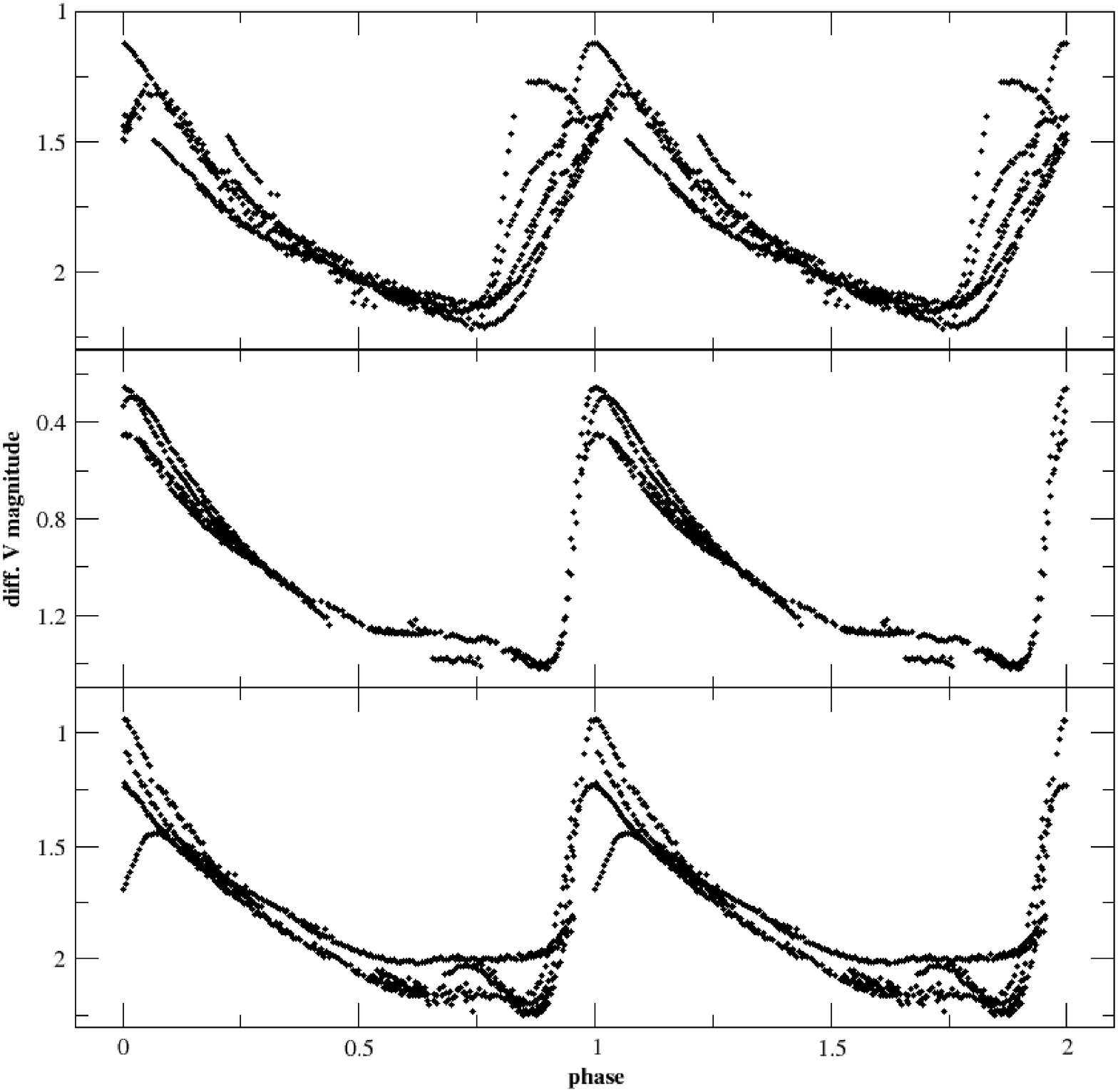}
\caption{RR~Lyr stars with Blazhko effect. From bottom to top: V7, V67, V68.}
\label{blazsko}
\end{center}
\end{figure}

It is known that a large fraction of RR~Lyr stars 
(20-30\% of RRabs and 2\% of RRcs, Kov\'acs 2001) show periodic amplitude 
and/or phase modulations, the so-called Blazhko effect, which is one
of the greatest mysteries in classical pulsating star research.
In our sample, three variables, V7, V67 and V68, have strong light curve
modulations, while a further 14 stars (V15, V23, V26, V27, V29, V35, V39, V40, V43, V55,
V58, V64, V78, V80) exhibited subtler changes during the 15 months of
observations. Furthermore, V33, V34 and V50 also showed detectable 
variations but these objects are in the very crowded core of the cluster 
and their photometry is of lower quality.

Assuming that these light curve modulations are representative for the whole sample, the
occurrence rate is rather high: about half of the cluster RR~Lyrae stars show some 
kind of changes in the phased data over the course of our monitoring. The three stars with
the strongest modulations are shown in Fig.\ \ref{blazsko}, where one can see that
changes in the light curve shape occurred in almost all phases. Very recently, Jurcsik
and co-workers obtained high-quality observations for a number of field RR~Lyraes
with the Blazhko effect (Jurcsik et al. 2006, S\'odor et al. 2006, Jurcsik et al.
2005),  revealing many puzzling features of the observed modulations, such as the
confinement of the light curve changes to a certain range of phases or the presence of
doubly-periodic modulations. Our data are not extensive enough to study 
these phenomena, but
it is obvious that there is a valuable set of RR~Lyraes in the cluster. 
With no firm theoretical explanation of the effect, a
globular cluster with many modulated RR~Lyraes clearly deserves further investigations.
Based on our observations, NGC~362 may be an excellent target for obtaining long-term,
homogeneous photometry of variables showing the Blazhko effect.

\section{The Color-Magnitude Diagram}

\begin{figure}
\begin{center}
\leavevmode
\includegraphics[width=8cm]{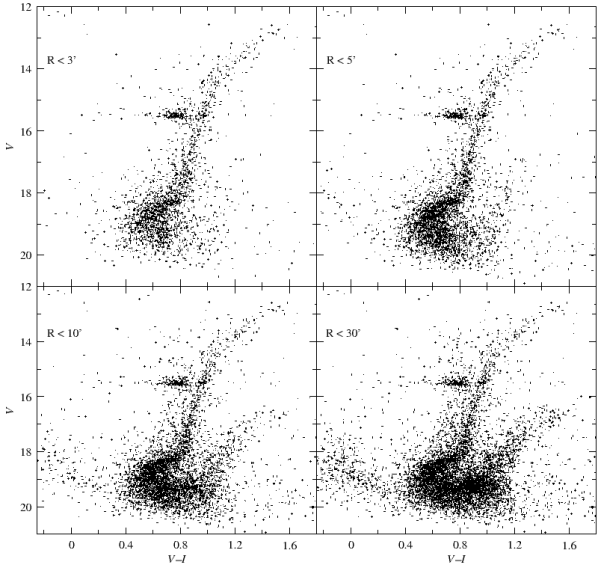}
\caption{$V, V-I$ CMDs with different star samples within increasing 
apertures centered on NGC~362. The aperture radii were 3, 5, 10 and 30 arc minutes. 
The outer regions clearly show a secondary faint RGB/AGB caused by stars in the SMC. 
The isochrones were fitted to data in the upper right ($R<5'$) panel.}
\label{cmds}
\end{center}
\end{figure}

Fig.\ \ref{cmds} shows CMDs for stars within an increasing 
area around the cluster. The radii were chosen as 3$^\prime$, 5$^\prime$, 10$^\prime$ 
and 30$^\prime$, leading to 2450, 4060, 5870 and 8567 stars in each plot.
As we increase the radius, the CMD exhibits an increasing field contamination and the
well-defined faint secondary Red Giant Branch (RGB) and Asymptotic Giant Branch (AGB)
of the SMC. We selected the second diagram 
(4060 stars) for isochrone fitting.

\begin{figure}
\begin{center}
\leavevmode
\includegraphics[width=8cm]{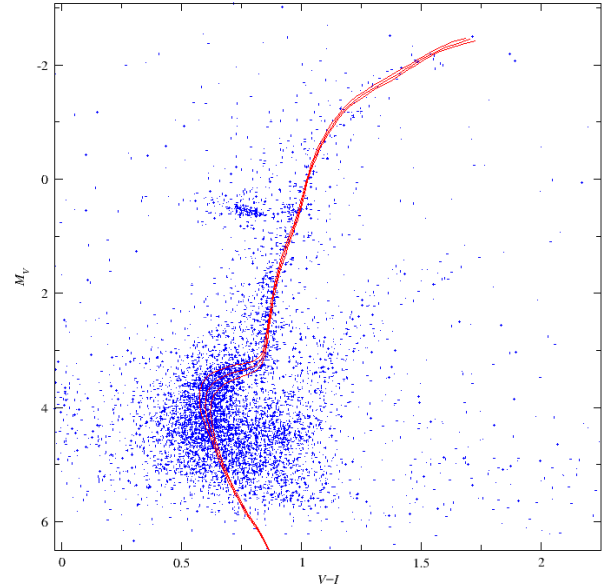}
\caption{$V, V-I$ CMD of the cluster with fitted $Z=0.001$ isochrones for 10, 11 and 
12 Gyr (from top). The derived  cluster parameters are $\mu=14.82\pm0.10$ and 
$E(V-I)=0.05$}
\label{fig.fit}
\end{center}
\end{figure}

To determine the distance modulus, age, metallicity and reddening of the cluster we
fitted Yonsei-Yale (Y$^2$) isochrones (Demarque et al., 2004).
The best fit was given by the 11 Gyr isochrone with $Z=0.001$,
$Y=0.232$, $[Fe/H]=-1.5$ and $[\alpha/Fe]=0.3$. It yields a distance modulus
of $\mu=14.95\pm0.1$ and reddening of $E(V-I)=0.05$. The two next best fits are also shown in
Fig.\ \ref{fig.fit},  which gave very similar distance and reddening estimates 
but have an age of 10 and 12 Gyr, respectively. These two fits bracket the 11 Gyr
isochrone in Fig.\ \ref{fig.fit}.

Between the HB and the RGB we also detect the Red Clump (RC) as a distinct concentration 
of stars at $V-I\approx0.96$ mag and $V\approx15.5$. From the intrinsic color of the RC
($V-I=0.92$, Olsen \& Salyk, 2002) and absolute brightness 
(for local Red Clump $M_{\rm V} =0.73$, Alves et al. 2002) we can
independently estimate the reddening and distance to the cluster: $E(V-I)=0.04$ and
$\mu=14.82$, both in good agreement with the results of the isochrone fitting. 

We corrected the distance modulus for extinction using $E(V-I)/E(B-V)=1.28$ 
(adopting $A_{\rm I}/E(B-V)=1.962$ and $A_{\rm V}/E(B-V)=3.24$, Savage \& Mathis 1979; 
Schlegel et al. 1998), thus $A_{\rm I}=0.08$. In this way we arrived at $E(B-V)=0.04$ and,
using $A_{\rm I}/A_{\rm V}=0.601$, we obtained $A_{\rm V}=0.13$, resulting in a true distance
modulus of $\mu_0=14.82\pm0.1$.

While the red HB is heavily populated, there is a noticeable lack of stars
in the blue wing of the HB (see a recent discussion of this feature by Bellazzini et
al.  2001 and Catelan et al. 2001). From Fig.\ \ref{fig.fit} we measured the
following   relative age horizontal and vertical parameters (Rosenberg et al. 1999):
$\delta ({\it V-I})_{2.5}=0.305\pm0.05$ and $\Delta {\it V}_{\rm TO}^{\rm
HB}=3.4\pm0.1$. Both values agree within the uncertainties with those measured by Rosenberg
et al. (1999). The brightness of the horizontal branch is $V_{\rm HB}=15.53\pm0.08$,
which is consistent with the RR~Lyr-based relation  
$M_{\rm V}(HB)=0.22[Fe/H]+0.89$ relation (Gratton et al. 2003), which predicts 
$V_{\rm HB}=15.58$.  We also calculated the
Horizontal Branch morphology parameter $L$, as defined by Lee (1990): $L=(B-R)/(B+R+V)$,
where $B$, $R$ and $V$ are the numbers of blue, red, and variable HB stars,
respectively. The result is $L=-0.855$, which can be  considered as an improvement to
the currently available value ($-0.87$, Harris 1996), thanks to our better statistics
of variable HB stars.

\section{Other short-period variable stars}

\begin{figure}
\begin{center}
\leavevmode
\includegraphics[width=8cm]{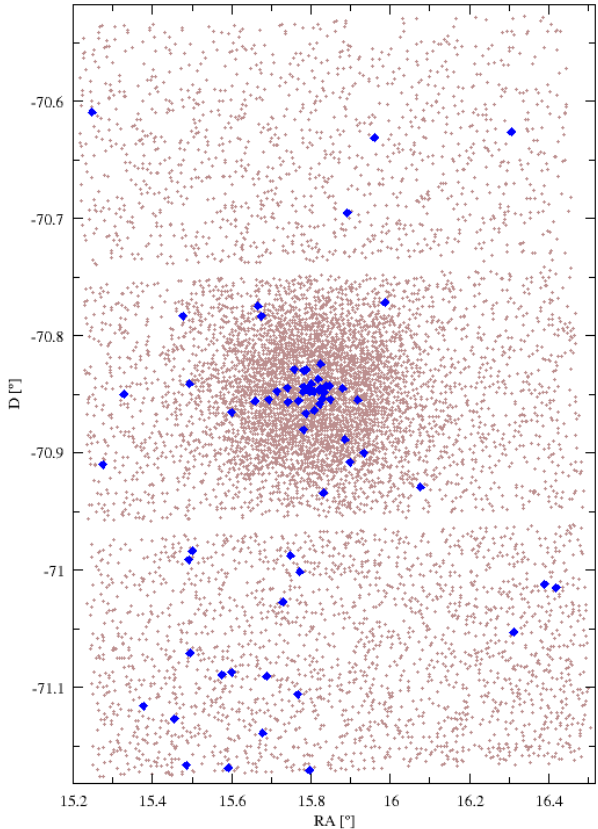}
\caption{RA--Dec positions of the variables stars projected on the positions of 
all the stars for which WCS transformation could be done. Blank sections are due to 
gaps between CCD segments. Note the increasing stellar density towards the SMC 
(in southern direction).}
\label{varcoo}
\end{center}
\end{figure}

We found a reasonably large number of variable stars other than RR~Lyraes
in the field of NGC~362. In Fig.\ \ref{varcoo} we plot the celestial positions 
of 8616 stars for which transformation to World Coordinate System could be achieved,
with all variable stars shown separately. Apparently, most of the cluster variables are
located in the central regions. The excess of
variables in the lower part of the figure is caused by the Small Magellanic
Cloud, located just outside of the field of view towards the south.

\begin{figure*}
\begin{center}
\leavevmode
\includegraphics[width=17cm]{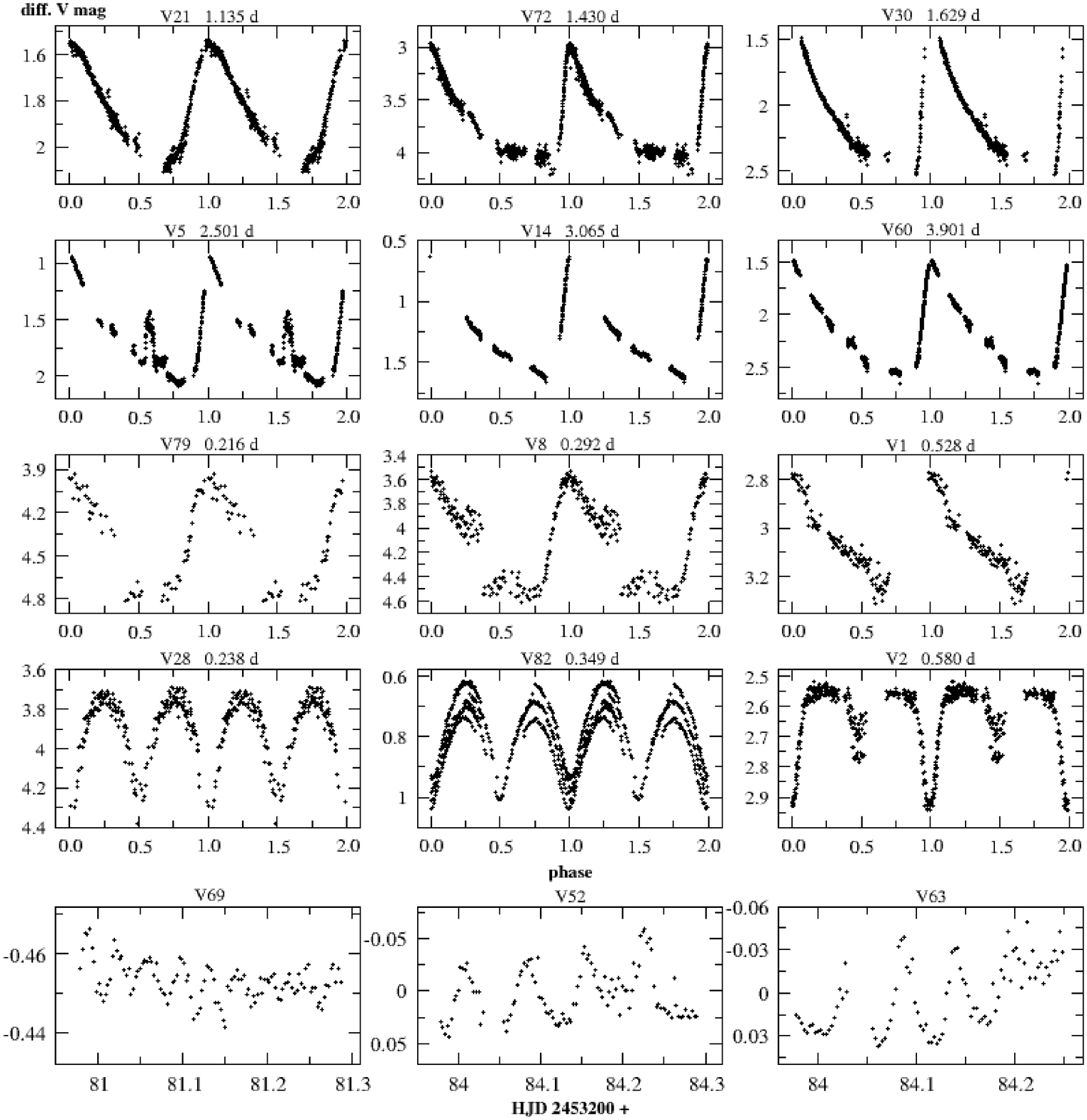}
\caption{Sample light curves of field variable stars. Top and second rows: Cepheids;
third row: eclipsing binaries, bottom row: $\delta$ Scuti stars}
\label{otherlcs}
\end{center}
\end{figure*}

Selected light curves are shown in  Fig.\ \ref{otherlcs}, while the basic
parameters are  listed in Table\ \ref{others1}. Here the $V-I$ colors are again only
indicative because of the single-epoch standard $I$ observation. $V$ brightnesses were
corrected for variability in the same way as for the RR~Lyraes. Besides eclipsing
binaries, field $\delta$ Scuti stars and long-period variables, the most interesting
objects include several Cepheids (three new discoveries) that can be identified  as
members of the Small Magellanic Cloud. Their positions in the sky fall closer to the
SMC, which strengthens the identification.

\begin{table*}
\begin{scriptsize}
\caption{Other variable stars in the field of {NGC 362}. The last column 
contains other designations of already known variables:  SAW xxx --- Sawyer (1931); 
SMC Vxxx --- Samus et al. (2004); SSP xx and HV xx --- Sharpee et al. (2002).}
\label{others1}
\centering
\begin{tabular} {ccclccccll}
\hline
\hline
\noalign{\smallskip}
Designation  & $\alpha$ [h:m:s] & $\delta$ [\r{}:$\arcmin$:$\arcsec$] & Period [d] &{\it V}& $V-I$& ampl. & Epoch & Type & Comments \\
&&&&&&[mag]&[HJD$-$2\,400\,000]&&\\
\hline
\noalign{\smallskip}
V2 & 1:00:59.28 & $-$70:36:33.4 & 0.58022 & 16.58 & 0.24 & 0.42 & 2926.0898 & EA & \\
V5 & 1:01:30.52 & $-$71:06:57.4 & 2.50181 & 16.51 & 0.63 & 1.16 & 2926.9921 & Cep & \\
V9 & 1:01:57.97 & $-$70:59:28.4 & $\sim$1.26 & 15.65 & $-$0.26 & 0.38 &  & EW & \\
V11 & 1:01:58.90 & $-$71:04:15.6 & 2.01329 & 15.83 & 0.41 & 0.66 & & Cep & SMC V1684, HV 1883 \\
V13 & 1:02:17.92 & $-$71:05:21.9 & 0.9624 & 18.49 & $-$0.01 & 0.58 & & EA & \\
V14 & 1:02:21.92 & $-$71:10:08.3 & 3.0650 & 16.28 & 0.69 & 1.03 & 2923.1352 & Cep & SMC V1709, HV 1906 \\
V21 & 1:02:42.53 & $-$71:08:21.6 & 1.13503 & 16.66 & 0.50 & 0.57 & 2921.1134 & Cep & SSP 32 \\
V22 & 1:02:45.24 & $-$71:05:26.2 & 1.59262 & 16.61 & 0.47 & 0.48 & 2850.1864 & Cep & \\
V28 & 1:02:59.45 & $-$70:59:16.2 & 0.23801 & 18.94 & 1.11 & 0.70 & 3284.1647 & EW & \\
V30 & 1:03:04.18 & $-$71:06:22.8 & 1.62998 & 17.03 & 0.59 & 1.05 & 2926.9921 & Cep & \\
V47 & 1:03:13.14 & $-$70:51:00.8 & & & & & & EA & \\
V51 & 1:03:15.19 & $-$70:50:36.7 & & & & & & SRA & SMC V1774 \\
V52 & 1:03:17.11 & $-$70:51:28.1 & & 14.68 & 0.72 & 0.10 & & DS/SX & \\
V56 & 1:03:17.63 & $-$70:50:49.4 & & 12.30 & $-$0.16 & & & EA & \\
V60 & 1:03:19.37 & $-$70:56:03.3 & 3.90134 & 15.86 & 0.44 & 1.12 & 2855.2316 & Cep & SAW V8 \\
V63 & 1:03:22.87 & $-$70:51:14.6 & & 15.51 & 0.70 & 0.09 & & DS/SX & \\
V69 & 1:03:34.09 & $-$70:41:42.8 & & 13.55 & 0.35 & 0.025 & & DS & \\
V72 & 1:03:44.13 & $-$70:54:01.1 & 1.43055 & 17.34 & 0.35 & 1.19 & 2929.1706 & Cep & SAW V15 \\
V75 & 1:04:10.11 & $-$70:57:21.5 & 4.205 & & & 0.66 & & Cep & SMC V1851, HV 214 \\
V79 & 1:05:15.74 & $-$70:37:35.7 & 0.21630 & 19.85 & 1.01 & 0.88 & 3281.0745 & HADS & halo/SMC \\
V82 & 1:05:38.75 & $-$70:44:09.7 & 0.34999 & & & 0.40 & 2921.1326 & EW & \\
\hline
\end{tabular}
\end{scriptsize}
\end{table*}

\begin{figure}
\begin{center}
\leavevmode
\includegraphics[width=8cm]{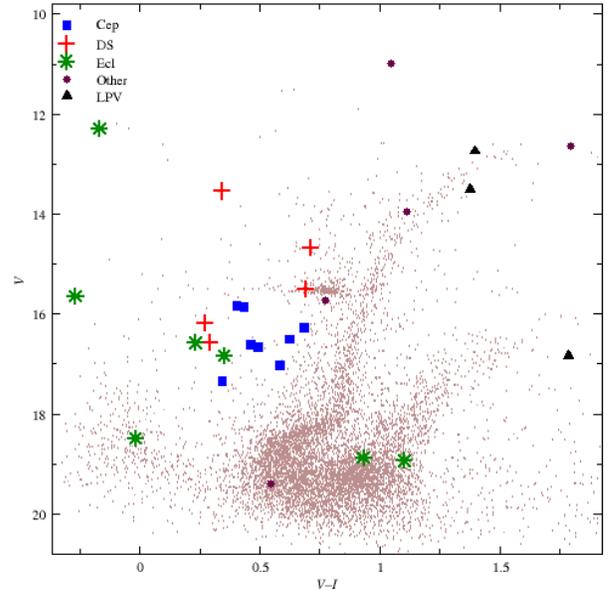}
\caption{Variable stars of other types than RR~Lyr in the color-magnitude 
diagram of NGC~362.}
\label{allcmd}
\end{center}
\end{figure}

In Fig.\ \ref{allcmd} we plotted the locations of non-RR~Lyr variables in the
CMD. Blue squares show the Cepheids, which indeed occupy the
expected instability strip of the SMC. Moreover, their mean $V$ magnitudes strictly
follow the Cepheid period--luminosity relation $V=-2.76\log P+17.61$, based
on OGLE observations of the SMC (Udalski et al. 1999). In addition, we can clearly
identify four cluster member red giant variables on the AGB and RGB and 
one in the SMC.

$\delta$~Sct-type variables show a large range in apparent magnitude, showing that they
belong to the galactic field. Based on its colour, magnitude and proximity to
NGC~362, V63 may be member of the cluster thus belonging to the SX~Phe class of variable
stars. Two of these stars, V63 and V69, exhibited complex multiperiodic variations,
from which we determined several pulsation frequencies with iterative sine-wave fitting
using Period04 of  Lenz \& Breger (2005). Although a thorough pulsation analysis of these
stars is beyond the scope of this paper, we list the resulting frequencies in Table\
\ref{dspars}. We included all frequencies with S/N ratio (Breger et al. 1993) higher than
4. For V63, we suspect that the low-frequency components can be caused by instrumental
effects and/or poor sampling. Otherwise, the frequency ratios of the strongest modes
suggest low-order radial pulsation, possibly flavoured with non-radial modes.

\begin{table}
\caption{Frequencies for two $\delta$~Scuti/SX~Phe stars.}
\label{dspars}
\centering
\begin{tabular} {lrr}
& V63 &\\
\hline
\noalign{\smallskip}
Frequency & cd$^{-1}$ & $V$ amplitude \\
\hline
\noalign{\smallskip}
$f_1$ & 17.1721 & 0.014 \\
$f_2$ & 5.26044 & 0.010 \\
$f_3$ & 21.4307 & 0.008 \\
$f_4$ & 6.00542 & 0.006 \\
$f_5$ & 11.8310 & 0.004 \\
$f_6$ & 17.8147 & 0.005 \\
\hline
\end{tabular}
\begin{list}{}{}
\item \hskip1cm S/N for $f_6$: 5.035\\
\end{list}
\begin{tabular} {lrr}
& V69 & \\
\hline
\noalign{\smallskip}
Frequency & cd$^{-1}$ & $V$ amplitude \\
\hline
\noalign{\smallskip}
$f_1$ & 27.6630 & 0.003 \\
$f_2$ & 33.0097 & 0.003 \\
$f_3$ & 28.7716 & 0.002 \\
$f_4$ & 35.8401 & 0.001 \\
\hline
\end{tabular}
\begin{list}{}{}
\item \hskip1cm S/N for $f_4$: 3.923\\
\end{list}
\end{table}

Finally, we found a number of variable stars, for which neither periods nor types
could be determined. They are listed in Table\ \ref{others2}.

\begin{table}
\begin{scriptsize}
\caption{Variable stars with no period and secure classification.}
\label{others2}
\begin{center}
\begin{tabular} {ccccccl}
\hline
\hline
\noalign{\smallskip}
Des. & $\alpha$ [h:m:s] & $\delta$ [\r{}:$\arcmin$:$\arcsec$] & $V$ & $V-I$ & ampl. & Notes \\
\hline
\noalign{\smallskip}
V3 & 1:01:06.09 & $-$70:54:35.5 & 16.18 & 0.28 & 0.1 & DS \\
V4 & 1:01:18.94 & $-$70:51:01.5 & 10.98 & 1.05 & & \\
V17 & 1:02:37.82 & $-$70:51:21.8 & 13.51 & 1.38 & & LPV \\
V18 & 1:02:37.93 & $-$70:50:42.3 & & & & EA \\
V19 & 1:02:39.48 & $-$70:46:28.9 & 16.83 & 1.79 & & LPV \\
V20 & 1:02:42.00 & $-$70:47:00.0 & 16.57 & 0.30 & 0.09 & DS \\
V24 & 1:02:51.13 & $-$70:50:50.3 & 13.94 & 1.12 & & \\
V25 & 1:02:54.85 & $-$71:01:39.3 & 18.23 & $-$0.25 & 0.8 & Cep?\\
V32 & 1:03:05.12 & $-$71:00:06.3 & 16.84 & 0.36 & 0.25 & EA \\
V36 & 1:03:07.84 & $-$70:49:46.6 & 12.64 & 1.40 & & LPV \\
V56 & 1:03:17.63 & $-$70:50:49.4 & 12.30 & $-$0.16 & & EA \\
V59 & 1:03:18.79 & $-$70:50:37.5 & & & & \\
V61 & 1:03:20.35 & $-$70:50:54.7 & 12.65 & 1.80 & & \\
V66 & 1:03:25.43 & $-$70:50:04.3 & & & & \\
V70 & 1:03:35.11 & $-$70:54:29.2 & 19.40 & 0.55 & & \\
V76 & 1:04:18.78 & $-$70:55:45.7 & 14.76 & 1.14 & & Cep?\\
V77 & 1:04:26.74 & $-$70:50:09.8 & & & & \\
V83 & 1:05:40.59 & $-$71:00:53.2 & 18.88 & 0.94 & 0.8 & EA \\
V84 & 1:06:19.92 & $-$71:04:55.3 & & & 0.12 & LPV \\
\hline
\end{tabular}
\end{center}
\end{scriptsize}
\end{table}

\section{Notes on some individual stars}

In this Section we give additional information for some of the detected 
variable stars. 
{\parskip-5mm

\paragraph{V1:} The given period in Sharpee et al. (2002, hereafter S02) 
is 1.16736 d, however, their phase diagram shows quite large scatter. 
Since our data define a much smoother phase curve, we prefer the period value
listed in Table\ \ref{rrls} (0.52835 d).

\paragraph{V4:} ISIS revealed the small-amplitude variations of this object only on
one night. 

\paragraph{V5:} This Cepheid variable had a peculiar bump on the descending branch
of its light curve. This phenomenon was detected only one night, possibly caused by
instrumental effects.

\paragraph{V7: } Based on our spectroscopic observations the mean radial velocity of
this star is 210 km~s$^{\rm -1}$.

\paragraph{V8:} The period in S02 is 0.57382d, however their light curve is
rather noisy. With full phase coverage our period (0.29218 d) seems to be the correct
one.

\paragraph{V10:} High-amplitude, continouos variation with a descending branch of
the flux curve, characteristic of an RR~Lyr star.

\paragraph{V11:} We confirm the period and classification by S02. 

\paragraph{V12:} We have only one night of data; the star is far from the 
cluster and is 4 magnitude fainter than the HB -- likely in the SMC.

\paragraph{V14:} We confirm the period and classification by S02. 

\paragraph{V17:} The star lies on the RGB of the cluster. The phased flux curve
suggests a period of 69 d. 

\paragraph{V19:} This star is located at the tip of the RGB of the SMC, 
possibly a long period red giant variable. The object is listed 
in the catalogue of carbon stars in the SMC compiled by 
Morgan \& Hatzidimitriou (1995).

\paragraph{V21:} We confirm the period and classification by S02.

\paragraph{V23:} Clement (2002) catalogued the object as a fundamental mode Cepheid
with 4.20519 d period. That is an error due to a mistyped decimal point, because the
star is an RR~Lyrae variable with 0.420234 d period (ten times smaller). Its mean
radial  velocity is 230 km~s$^{\rm -1}$.

\paragraph{V31:} The star showed characteristic RR~Lyr-like light curves but also a
strong modulation from day to day. Both the extrema and the amplitude varied  on a
short time-scale, thus our efforts to determine the pulsational parameters  were not
successful. After checking an archived HST image, we found the star to be a very close
blend. The constituting stars are separated only by 159 mas from each other and seem
to have the same brightness. Therefore, we concluded that both components are
RR~Lyrae stars and the superposition of their individual light curves caused the
failure of the period determination. Their angular separation corresponds to 1200 AU
at the distance of the cluster, so they may be physically related objects.

\paragraph{V36:} It is included in the Catalogue of $[Fe/H]$ values for  F, G, K
stars (Cayrel de Strobel et al. 2001) with $[Fe/H]=-1.33$, in
accordance with the metallicity of the cluster.

\paragraph{V38:} This star is on the HB, while one night of ISIS
data showed an ascending branch of the flux curve.

\paragraph{V41:} This star is on the HB with an amplitude larger than 0.5 mag.

\paragraph{V47:} This star is listed in the catalog of Galactic
Globular Clusters (Monella 1985) with the following parameters: $[Fe/H]=-1.39$, 
$v_{\rm rad}$=--221 km~s$^{\rm -1}$, clearly belonging to the cluster. 
The light curve is characteristic of an Algol-type eclipsing binary with hints 
of small amplitude pulsations in maximum. Unfortunately, it is located in a very
crowded field so that the light curve is often noisy. 

\paragraph{V48:} The one night of ISIS data shows high-amplitude sinusoidal light
variation. The star is located in the center of the cluster and is possibly an 
RRc star.

\paragraph{V50:} The phased ISIS flux data show strong light curve
variations; problems with crowding in the cluster center. 

\paragraph{V51:} Samus et al. (2004) gives a period of 135 d. 

\paragraph{V60:} We confirm the results by Sawyer (1931).

\paragraph{V61:} This object is located at the tip of the RGB of the cluster, 
probably a long-period red giant variable.

\paragraph{V62:} The star is on the HB, suggesting that it is
an RR~Lyr variable. The short flux curve shows an ascending branch. 

\paragraph{V63:} Although the star seems to be on the HB, its light curve is 
more characteristic of a $\delta$ Sct-type variable. 

\paragraph{V74:} The mean radial velocity of the star is 230 km~s$^{\rm -1}$.

\paragraph{V75:} This star was in the gap between the CCD segments at the time of
standard observation, that is why there are no $V$ and $V-I$ values in Table\
\ref{others1}. The light curve confirms the period and classification by S02.

\paragraph{V82:} This W~UMa-type variable has interesting changes in the 
mean level of the light curve. We could not find any source of systematical
errors in our data reductions, so the behaviour seems to be real 
(see also Walter 1983 discussing the so-called Kwee-effect) 
}

\section{Conclusions}

The main results of this paper can be summarized as follows.

\begin{itemize}

\item Using PSF and image subtraction photometry we found 45 RR Lyr variables in the
field of the cluster, of which 28 are new discoveries. We converted flux curves into
magnitudes for all RR~Lyrae stars with a simple method, which allowed us to use
empirical light curve shape vs. physical parameter calibrations. With these, we
determined metallicities, absolute magnitudes, reddenings and other physical parameters
for 16 RR~Lyraes. 

\item We found a rather high-percentage of modulated RR~Lyr light curves, i.e. the
Blazhko effect, both for RRab and RRc-type stars. NGC~362 could be an excellent target
for studying the Blazhko effect in a chemically homogeneous environment. 

\item From isochrone fitting we also determined the main parameters of the cluster.
Both the RR~Lyraes and the color-magnitude diagram result in a consistent set of
physical parameters.

\item We also discovered variable stars of other types, including Cepheids and
long-period variables in the Small Magellanic Cloud, eclipsing binaries and
$\delta$~Sct-type pulsating stars in the galactic field. 

\end{itemize}

\begin{table}
\caption{Physical parameters of NGC~362.}
\setlength{\tabcolsep}{1mm}
\label{table.comp}
\centering
\begin{tabular}{llll}
\hline
\hline
\noalign{\smallskip}
Reference & $[Fe/H]$ & Distance [kpc] & Age [Gyr] \\
\hline
\noalign{\smallskip}
{\it This paper:} & & & \\
RR~Lyraes & $-$1.16$\pm$0.25 & 7.9$\pm$0.6 & -- \\
{\it VI} CMD & & 9.2$\pm$0.4 & 10--12 \\
{\it Previous studies:} & & & \\
RS05 & $-$1.22 & -- & 9.1 \\
MG04 & $-$1.16 & --  & -- \\
DA05 & $-$1.33 & 9.3 & 8.4 \\
SK00 & $-$1.33 & -- & -- \\
FW93 & $-$1.03 & 9 & 13 \\
ZW84 & $-$1.27 & -- & -- \\
HR96 & $-$1.16 & 8.5 & -- \\
\hline
\end{tabular}
\begin{list}{}{}
\item RS05 - Rakos \& Schombert (2005)\\
\item MG04 - Mackey \& Gilmore (2004)\\
\item DA05 - De Angeli et al. (2005)\\
\item SK00 - Shetrone \& Keane (2000)\\
\item FW93 - Fischer et al. (1993)\\
\item ZW84 - Zinn \& West (1984)\\
\item HR96 - Harris (1996)\\
\end{list}
\end{table}

A brief overview of the main physical parameters of NGC~362 is given  in Table\
\ref{table.comp}.  Of these, the only significant difference is between  the distance
estimates using the RR~Lyraes and the color-magnitude diagram. The difference is likely to
originate predominantly from the RR~Lyraes, whose absolute magnitude--metallicity
relation is not yet settled, with a number of suggestions about nonlinear
calibrations of $M_{\rm V}/[Fe/H]$ (see Sect.\ 5 of Sandage \& Tamman 2006).

\begin{acknowledgement}

This project has been supported by the Hungarian OTKA Grant
\#T042509, a Hungarian E\"otv\"os Fellowship to PSz and the Australian Research 
Council. LLK is supported by a University of Sydney Postdoctoral Research 
Fellowship. The NASA ADS Abstract  Service was used to access references. This 
work used SIMBAD and Vizier databases operated at CDS-Strasbourg, France.

\end{acknowledgement}

{}
\end{document}